\documentclass[preprint,showpacs,eqsecnum,aps,amsmath,amssymb,floatfix]{revtex4}
\usepackage{graphics}
\usepackage{epsfig}
\usepackage{graphicx}
\usepackage[bookmarks=false]{hyperref}
\usepackage{ifpdf}
\begin{document}

\title{Towards a new semi-classical interpretation of the wobbling motion in $^{163}$Lu}

\author{A. A. Raduta$^{a), b)}$,  R. Poenaru $^{a),c)}$ and C. M. Raduta $^{a)}$ }

\affiliation{$^{a)}$ Department of Theoretical Physics, Institute of Physics and
  Nuclear Engineering, Bucharest, POBox MG6, Romania}

\affiliation{$^{b)}$Academy of Romanian Scientists, 54 Splaiul Independentei, Bucharest 050094, Romania}

\affiliation{$^{c)}$Doctoral School of Physics, Bucharest University, 405 Atomistilor Str., Bucharest-Magurele, Romania}

\begin{abstract}
 A new interpretation for the wobbling bands in $^{163}$Lu is given within a particle-triaxial rotor semi-classical formalism. While in the previous papers the  bands TSD1, TSD2, TSD3 and TSD4 are viewed as the  ground, one, two and three  phonon wobbling bands, here the corresponding experimental results are described as the ground band with  spin equal to I=R+j, for R=0,2,4,...(TSD1), the  ground band with I=R+j and R=1,3,5,...(TSD2), the one phonon excitations of TSD2 (TSD3), with the odd proton moving in the orbit  $j=i_{13/2}$, and the ground band of I=R+j, with R=1,3,5,... and $j=h_{9/2}$ (TSD4). The moments of inertia (MoI) of the core for the first three bands are the same, and considered to be free parameters. Due to the core polarization effect caused by the particle-core coupling, the MoI's for TSD4 are different. The energies and the e.m. transitions are quantitatively well described. Also, the phase diagram of the odd system is drawn. In the parameter space one indicates where the point associated with the fitted parameters is located and also which is the region of transversal wobbling mode as well as where the wobbling motion is forbidden.
\end{abstract} 
\maketitle

\renewcommand{\theequation}{\arabic{equation}}
\setcounter{equation}{0}

The wobbling motion consists in a precession of the total angular momentum of a triaxial system combined with an oscillation of its projection on the quantization axis around a steady position. Bohr and Mottelson described the wobbling motion within a triaxial rotor  model for high spin states, where the total angular momentum almost aligns to the principal axis with the largest moment of inertia \cite{BMott}. This pioneering paper was followed by a fully microscopic description due to Marshalek \cite{Marsh}. Since then a large volume of experimental and theoretical results has been accumulated
\cite{Odeg,Jens,Ikuko,Scho,Amro,Gorg,Ham,Matsu,Ham1,Jens1,Hage,Tana3,Bring,Hart,MikIans,Rad016,Shi,Chen,Wu,Sensha}. Also, the concept of wobbling motion has been extended to even-odd nuclei. Experimentally, the wobbling states excited in triaxial strongly deformed (TSD) bands are known in several nuclei like $^{161,163,165,167}$Lu, $^{167}$Ta \cite{Bring,Hart}, $^{135}$Pr  
\cite{Matta,Tan017,Frau,Buda} and $^{187}$Au \cite{Sensha}.

In various versions, the theoretical phenomenological studies are based  on semi-classical descriptions. Thus, the equations of motion for the classical rotor Hamiltonian are exactly treated in Ref.
\cite{Rad98}, while in Refs.\cite{Ham,Frau} the harmonic approximation is adopted for the wobbling frequency. The approximation is justified for large angular momentum but not for values close to that of the band head state. Moreover, the odd particle angular momentum is rigidly coupled to the core, along the axis 1, of largest moment of inertia. In Ref.\cite{Rad017,Rad018}  the classical picture is obtained via a time dependent variational principle, the collective and individual coordinates being treated on  equal footing. The quantal treatment uses the boson description of the angular momenta describing the even/odd system. Thus, the Holstein-Primakoff \cite{Tana3} and Dyson\cite{Oi} boson expansion methods have been used.  The drawback of such approaches consists in that the zero point energy is crudely approximated.
In Ref.\cite {Rad98} a new boson representation, in terms of elliptic functions, for the components for angular momenta is proposed. The Bargmann representation for the rotor Hamiltonian allows to separate the potential energy which provides an exact description for the ground state of the wobbling motion. Alternatively \cite{Chen}, such a separation is achieved by making use of the Pauli quantization recipe \cite{Pauli}. The microscopic theories use the random phase approximation (RPA) plus cranking, but, however, the higher RPA effects are ignored. 

Depending on the relative position of the rotation axis of the collective core and  that of the odd nucleon, the wobbling motion has a longitudinal or a transversal character. In the first case the two rotation axes are parallel, while for the transversal wobbling the rotation axis of the rotor and that of the odd nucleon are prependicular. In the latter case the particle-core interaction drives the whole system to a shape of a large and stable deformation, and this rotates around the axis of maximal moment of inertia. The concept of the transversal wobbling was introduced by Frauendorf in Ref. \cite{Frau}, but not confirmed by the calculations of Tanabe \cite{Tan017}. A comment about the debate on this issue is presented in the present paper. in the context of the phase diagram.

In a previous publication \cite{Rad017} we formulated a semi-classical formalism so as to describe the main features of the wobbling motion for a particle-triaxial-rotor system, which was successfully used for $^{163}$Lu. The odd particle is a proton in the $j=i_{13/2}$ orbital. The method was subsequently applied  to $^{165,167}$Lu \cite{Rad018}. Therein, each state of the TSD1 band is determined by a time dependent variational principle  equation under the restriction of small amplitudes. The solution leads to a phonon operator which applied successively to the ground states with the spin I=R+j and R=0,2,4,..., gives rise to the so called TSD2 band. Applying the phonon operator twice  to the TSD1 states, one obtains the TSD3 band. The states of the TSD4 have negative parity and are obtained by acting with three phonons, two of positive and one of negative parity, on the TSD1 states. The negative parity wobbling phonon corresponds to a $j=h_{9/2}$ proton coupled to a triaxial rotor with the moments of inertia modified due to the coupling of a new proton orbit. The phonon operator increases the spin of state by one unit. Also, the e.m. properties of the mentioned isotopes have been well described. The sketched approach is consistent with the experimental result claiming that it provides evidence of multiple wobbling phonon  states.

Here we present an approach which does not use the multi-phonon states in order to account for the experimental features of $^{163}$Lu. We begin with a brief description of the new method.

We thus study an odd-mass system  consisting of an even-even core described by a triaxial rotor Hamiltonian $H_{rot}$ and a single j-shell proton moving in a quadrapole deformed mean-field:
\begin{equation}
H_{sp}=\epsilon_j+\frac{V}{j(j+1)}\left[\cos\gamma(3j_3^2-{\bf j}^2)-\sqrt{3}\sin\gamma(j_1^2-j_2^2)\right].
\label{hassp}
\end{equation}
Here $\epsilon_j$ is the single particle energy and $\gamma$, the deviation from the axial symmetric picture.
In terms of the total angular momentum ${\bf I}(={\bf R}+{\bf j}) $ and the angular momentum carried by the odd particle, ${\bf j}$, the rotor Hamiltonian is written as:
\begin{equation}
H_{rot}=\sum_{k=1,2,3}A_k(I_k-j_k)^2.
\end{equation}
where $A_k$ are half of the reciprocal moments of inertia associated to the principal axes of the inertia ellipsoid, i.e. $A_k=1/(2{\cal I}_k)$, which are considered as free parameters. 

The expressions for the single particle coupling potential, $H_{sp}$, and the triaxial rotor term, $H_{rot}$, have been previously used by many authors, the first being Davydov \cite{Davy1,Davy2}.

The eigenvalues of interest for $\hat{H}(=H_{rot}+H_{sp})$ are obtained through a time dependent variational principle equations.
Thus, the total Hamiltonian $\hat{H}$ is dequantized through  the time dependent variational principle:
\begin{equation}
\delta\int_{0}^{t}\langle \Psi_{IjM}|{\hat H}-i\frac{\partial}{\partial t'}|\Psi_{IjM}\rangle d t'=0,
\end{equation}
with the trial function chosen as:
\begin{equation}
|\Psi_{Ij;M}\rangle ={\bf N}e^{z\hat{I}_-}e^{s\hat{j}_-}|IMI\rangle |jj\rangle ,
\end{equation} 
with $\hat{I}_-$ and $\hat{j}_-$ denoting the lowering operators for the intrinsic angular momenta ${\bf I}$ and ${\bf j}$ respectively, while ${\bf N}$ is  the normalization factor.
$|IMI\rangle $ and $|jj\rangle$ are extremal states for the operators ${\hat I}^2, {\hat I}_3$ and ${\hat j}^2, {\hat j}_3$, respectively. Note that the trial function is a linear combination of  components of definite K, which is consistent with the fact that for triaxial nuclei, $K$ is not a good quantum number. Some authors refer to the TSD bands as to the super-deformed bands suggesting that the ground band head state is an isomeric state with a relative large half-life.
 
The variables $z$ and $s$ are complex functions of time and play the role of classical phase space coordinates describing the motion of the core and the odd particle, respectively:
\begin{equation}
z=\rho e^{i\varphi},\;\;s=fe^{i\psi}.
\end{equation}
Changing the variables $\rho$ and $f$ to  $ r$ and $t$, respectively:
\begin{equation}
r=\frac{2I}{1+\rho^2},\;\;0\le r\le 2I;\;\;
t=\frac{2j}{1+f^2},\;\; 0\le t\le 2j,
\end{equation}
the classical equations of motion acquire the canonical form:
\begin{eqnarray}
\frac{\partial {\cal H}}{\partial r}&=&\stackrel{\bullet}{\varphi};\;\frac{\partial {\cal H}}{\partial \varphi}=-\stackrel{\bullet}{r} \nonumber\\ 
\frac{\partial {\cal H}}{\partial t}&=&\stackrel{\bullet}{\psi};\;\frac{\partial {\cal H}}{\partial \psi}=-\stackrel{\bullet}{t}. 
\label{eqmot}
\end{eqnarray}
where ${\cal H}$ denotes the average of $\hat{H}$ with the trial function $|\Psi_{IjM}\rangle$ and plays the role of the classical energy function.
The classical energy has the expression : 
\begin{eqnarray}
{\cal H}&\equiv&\langle \Psi_{IjM}|H|\Psi_{IjM}\rangle\nonumber\\
        &=&\frac{I}{2}(A_1+A_2)+A_3I^2+\frac{2I-1}{2I}r(2I-r)\left(A_1\cos^2\varphi+A_2\sin^2\varphi -A_3\right)\nonumber\\
        &+&\frac{j}{2}(A_1+A_2)+A_3j^2+\frac{2j-1}{2j}t(2j-t)\left(A_1\cos^2\psi+A_2\sin^2\psi -A_3\right)\nonumber\\
        &-&2\sqrt{r(2I-r)t(2j-t)}\left(A_1\cos\varphi\cos\psi+A_2\sin\varphi\sin\psi\right)+A_3\left(r(2j-t)+t(2I-r)\right)-2A_3Ij\nonumber\\
        &+&V\frac{2j-1}{j+1}\left[\cos\gamma-\frac{t(2j-t)}{2j^2}\sqrt{3}\left(\sqrt{3}\cos\gamma +\sin\gamma\cos2\psi\right)\right]
\label{classen}
\end{eqnarray} 
and is minimal (${\cal H}_{I,min}(j)$) in the point
$(\varphi,r)=(0,I);(\psi,t)=(0,j)$, when $A_1<A_2<A_3$.  
Linearizing the equations of motion around the minimum point of ${\cal H}$, one obtains a harmonic motion for the system, with the frequency given by the equation:
\begin{equation}
\Omega^4+B\Omega^2+C=0,
\label{ecOm}
\end{equation}
where the coefficients B and C have the expressions:
\begin{eqnarray}
-B&=&\left[(2I-1)(A_3-A_1)+2jA_1\right]\left[(2I-1)(A_2-A_1)+2jA_1\right]+8A_2A_3Ij\nonumber\\
 &+&\left[(2j-1)(A_3-A_1)+2IA_1+V\frac{2j-1}{j(j+1)}\sqrt{3}(\sqrt{3}\cos\gamma+\sin\gamma)\right]\nonumber\\
 &\times&\left[(2j-1)(A_2-A_1)+2IA_1+V\frac{2j-1}{j(j+1)}2\sqrt{3}\sin\gamma\right],\\
C&=&\left\{\left[(2I-1)(A_3-A_1)+2jA_1\right]\left[(2j-1)(A_3-A_1)+2IA_1+V\frac{2j-1}{j(j+1)}\sqrt{3}(\sqrt{3}\cos\gamma+\sin\gamma)\right]\right. \nonumber\\
&-&\left. 4IjA_3^2\right \}\nonumber\\
 &\times&\left\{\left[(2I-1)(A_2-A_1)+2jA_1\right]\left[(2j-1)(A_2-A_1)+2IA_1+V\frac{2j-1}{j(j+1)}2\sqrt{3}\sin\gamma\right]-4IjA_2^2\right\}.\nonumber\\
\label{BandC}
\end{eqnarray}
As mentioned in Ref.\cite{Rad018} in the expression for the classical energy of Ref.\cite{Rad017} in the fourth line of  Eq. (\ref{classen}) the factor 2  in the first term, which couples the variable $\varphi$, $\psi$ and the free term, $-2IjA_2^2$, are missing due a lamentable error. We checked that the same spectrum is obtained by a proper renormalization of the MoI's. 

Under certain restrictions for MoI's the dispersion equation (\ref{ecOm}) admits two real and positive solutions. Hear after these will be denoted by $\Omega^{I}_1$ and $\Omega^{I}_{1'}$ for $j=i_{13/2}$
and $\Omega_2$ and $\Omega_{2'}$ for $j=h_{9/2}$. These energies are ordered as: $\Omega^I_1<\Omega^I_{1'}$ and $\Omega^I_2<\Omega^I_{2'}$. Energies of the states in the four bands are defined as:
\begin{eqnarray}
E^{TSD1}_I&=&\epsilon_{13/2}+{\cal H}_{I,min}(13/2)+\frac{1}{2}\left(\Omega^I_1+\Omega^I_{1'}\right),I=13/2, 17/2, 21/2,.....\nonumber\\
E^{TSD2}_I&=&\epsilon_{13/2}+{\cal H}_{I,min}(13/2)+\frac{1}{2}\left(\Omega^I_1+\Omega^I_{1'}\right),I=27/2, 31/2, 35/2,.....\nonumber\\
E^{TSD3}_I&=&\epsilon_{13/2}+{\cal H}_{I-1,min}(13/2)+\frac{1}{2}\left(3\Omega^{I-1}_1+\Omega^{I-1}_{1'}\right),I=33/2, 37/2, 41/2,.....\nonumber\\
E^{TSD4}_I&=&\epsilon_{9/2}+{\cal H}_{I,min}(9/2)+\frac{1}{2}\left(\Omega^I_2+\Omega^I_{2'}\right),I=47/2, 51/2, 55/2,.....
\label{ener}
\end{eqnarray}
The spin sequences from Eq.(\ref{ener}) correspond to the  rule presented before, that is j+R with R=2n, n=0,1,2,... for TSD1, R=2n+1, n=0,1,2,...for TSD2 and R=2n, with n=1,2,3,... for TSD3 where R stands for the core angular momentum, while j is the angular momentum of the odd proton.
The excitation energies are obtained by subtracting $E^{TSD1}_{13/2}$ from the above expressions. Fitting the experimental data for the excitation energies,through a least square procedure one obtains the moments of inertia and the strength of the particle-core coupling potential. The results of the fitting procedure are collected in Table I.
\begin{table}
\begin{tabular}{|c|c|c|c|c|c|c|}
\hline
j& bands& ${\cal I}_1[\hbar^2/MeV]$& ${\cal I}_2[\hbar^2/MeV]$ &${\cal I}_3[\hbar^2/MeV]$& V[MeV] &$\gamma $ [degrees]\\
\hline
13/2&TSD1,TSD2,TSD3&63.2  &  20  &  10  & 3.1 &  17\\
9/2&TSD4           &67    &  34.5 & 50  & 0.7&  17\\
\hline
\end{tabular}
{\scriptsize
\caption{The MoI's, the strength of the single particle potential (V) and  the triaxial parameter ($\gamma$) as provided by the adopted fitting procedure.}}
\label{Table 1}
\end{table}
With the MoI's, V and $\gamma$ determined, Eqs.(\ref{ener}) give the energies for the four bands. The excitation energies are compared with the corresponding experimental data in Fig.1, where a good agreement can be seen. Note that for the first three bands the excitation energies do not depend on the single particle energies. On the contrary,  the  excitation energies for the TSD4 states contain the constant term $\epsilon_{9/2}-\epsilon_{13/2}=-0.334 MeV$, which according to Eqs. (\ref{hassp}) and (\ref{ener}) are just the difference in energy for the spherical shell model states $h_{9/2}$ and $i_{13/2}$. 

It is worth adding a short comment about the band parity. The states of the collective core can be classified by the irreducible representations of the discrete group of transformations $D_2$; these are characterized by the vectors $(r_0,r_1,r_2,r_3)$ where $r_k$ denotes the eigenvalues of the unity rotation ($r_0$) and  of the rotations with $\pi$ around the principal axes $x_k$ ($r_k$)with k=1,2,3, respectively. The states (1,1,1,1) and (1,-1,1,-1) are of positive parity while (1,-1,-1,1) and (1,1,-1,-1)
of negative parity. The mentioned representations are conventionally called $ A, B_2, B_1$ and $B_3$, respectively. Thus, the states of type $A$ and $B_2$ are of positive while those of the $B_1$ and $B_3$ kind are of negative parity. The parity of the odd system is obviously given by the product of parities brought by the core and single particle, respectively. In the case of TSD4 the factors of the parity product correspond to the states of the type $A$ or $B_2$ and the single particle orbit $h_{9/2}$. Microscopically, the particle-hole excitations involve one single particle state belonging to the core and the negative parity-odd particle.

 We notice that the least square fit predicts that the maximal moment of inertia corresponds to the one-axis and therefore the system rotates around the short axis. Moreover, the odd proton angular momentum is oriented also along the short axis and thereby the system motion is of longitudinal wobbling character. The numerical values of MoI's are consistent with the angular momenta orientation corresponding to the minimum point of ${\cal H}$. However, although our results agree well with the data they do not reproduce the decreasing behavior of the energy spacings with increasing angular momentum which, as claimed in Ref.\cite{Frau}, is a signature for a transversal wobbling (see Fig.2). This behavior is predicted by the microscopic calculation using a  cranking type of Hamiltonian. However, the conclusions of such calculations are induced by two main ingredients: i) the single particle basis are provided by a deformed mean field consistent with a hydrodynamic model, which favors the rotation around the middle axis; ii) The cranking term, which cranks the system to rotate around the short axis. The balance of the two effects determines one regime or another. It is worth mentioning that besides the very good agreement between predictions and experimental  data we compared, with a positive result, the energies of the four TSD bands with the exact eigenvalues of the starting Hamiltonian \cite{Rad018}. This confirms that the proposed formalism is  appropriate not only for simulating the data, but also  provides a good approximation for the exact results. In our opinion, the decreasing behavior of the wobbling energy with the spin is not a decisive test for a given formalism. As a matter of fact, using the standard definition, the wobbling energy  for the one phonon band, i.e.,the TSD3, was plotted in Fig. 2 as function of the angular momentum. Surprisingly, the wobbling frequency  increases slightly with spin, as predicted by our approach. Indeed, the experimental wobbling energy increases from 144 to 170 keV when the spin goes from $33/2$ to $77/2$ and finally decreases for the last two states, with spins 81/2 and 85/2, to 143 keV. On the other hand, the calculated wobbling energy increases faster with angular momentum, from 331 keV at spin 
$33/2$ up to 570 keV for spin $85/2$. The agreement between the wobbling energy behavior given by our calculations and the corresponding experimental data is to be considered as a specific feature of the present approach.
\begin{figure}[ht!]
\includegraphics[width=0.4\textwidth]{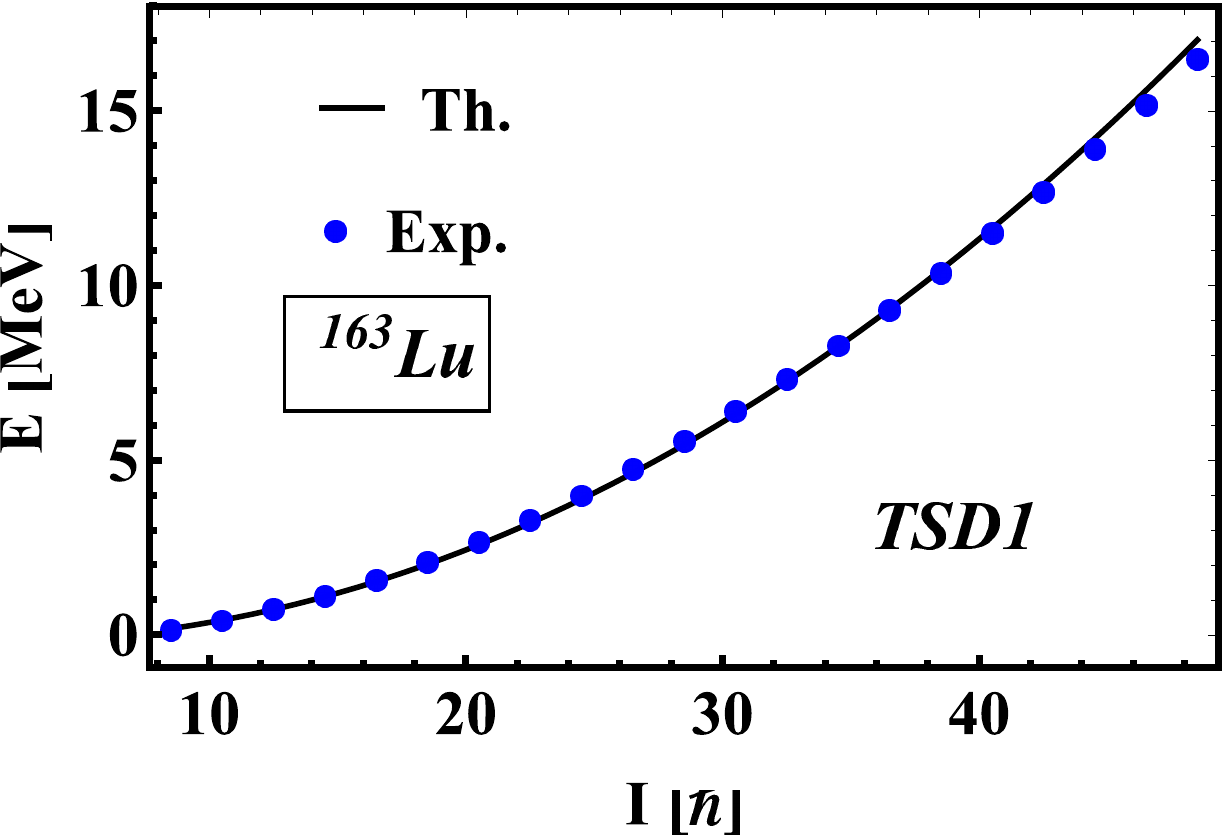}\hspace{0.1cm}\includegraphics[width=0.4\textwidth]{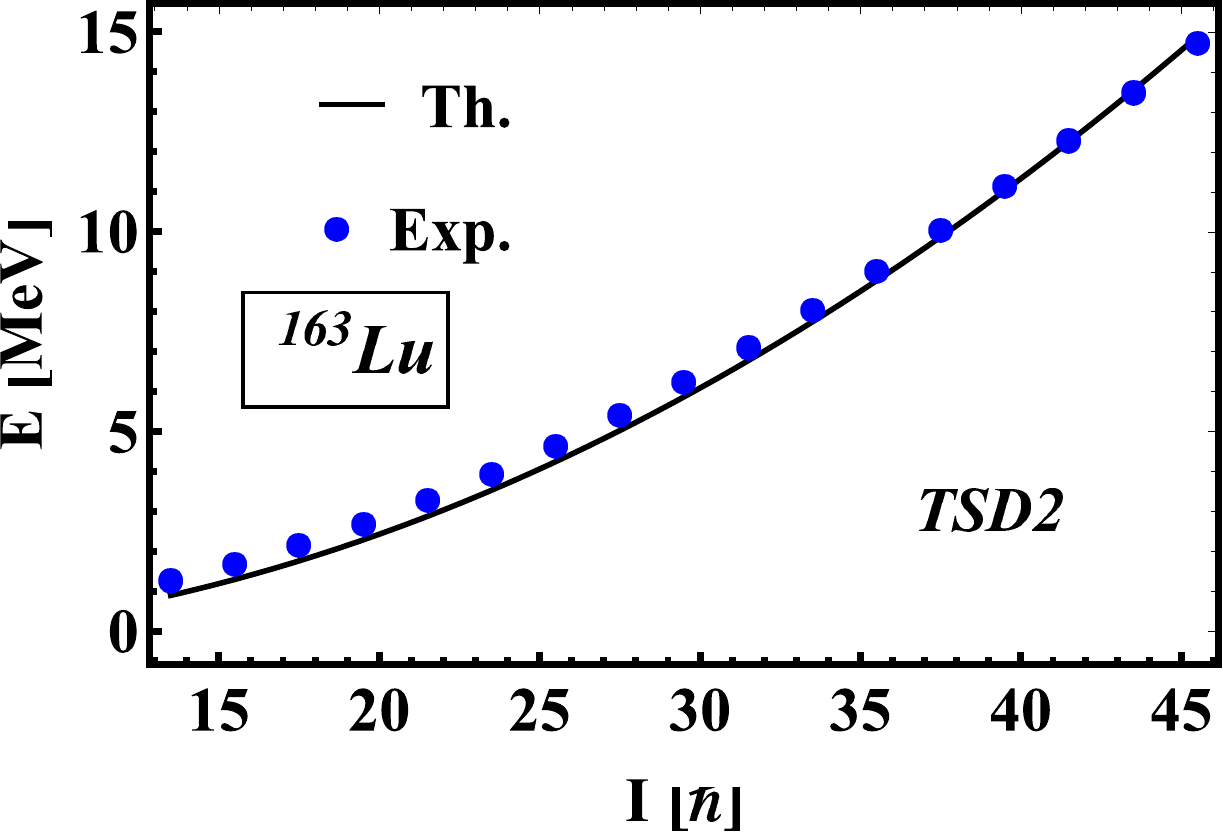}
\includegraphics[width=0.4\textwidth]{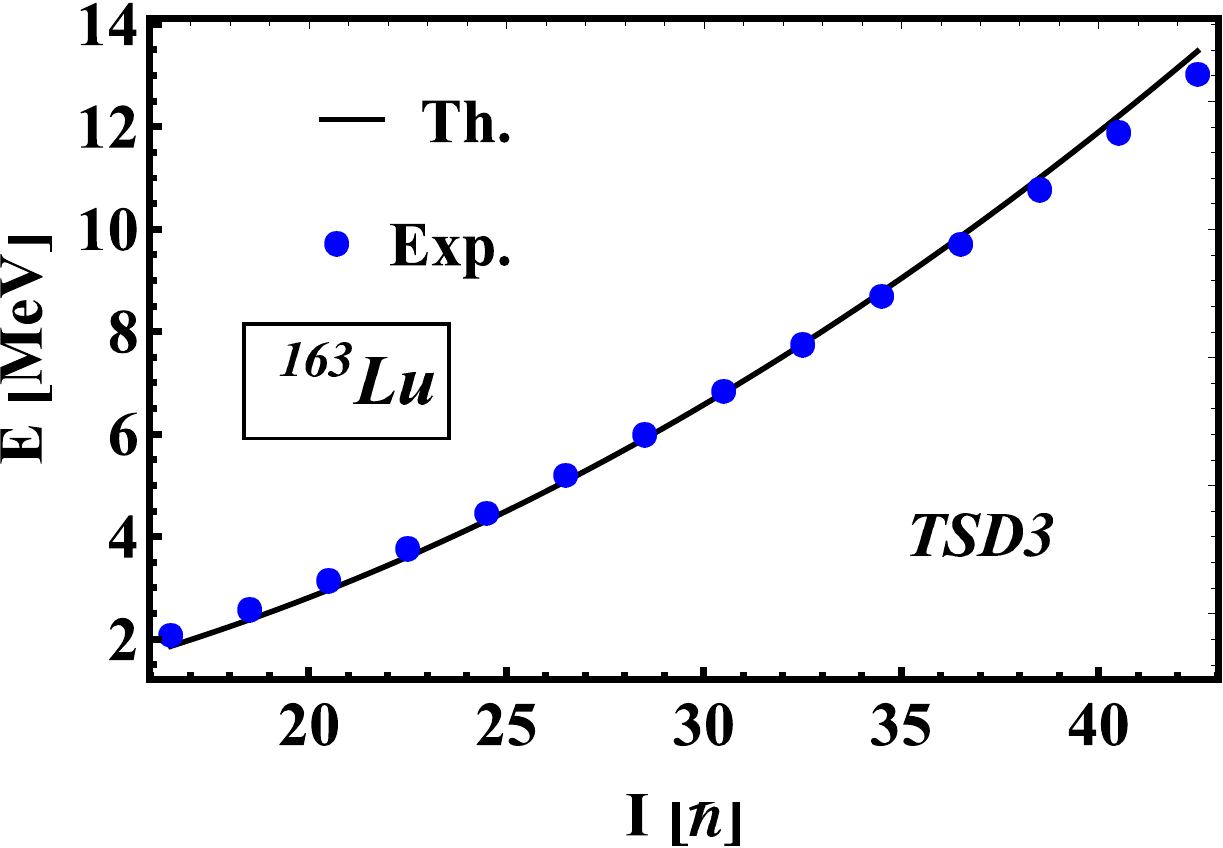}\hspace{0.1cm}\includegraphics[width=0.4\textwidth]{tsd3_en_163.pdf}
{\scriptsize
\caption{(Color online)Calculated energies for the bands TSD1, TSD2 and TSD3 are compared with the corresponding experimental data taken from Ref.\cite{Jens1}.}}
\label{Fig.1}
\end{figure}

\begin{figure}[ht!]
\includegraphics[width=0.25\textwidth,height=4cm]{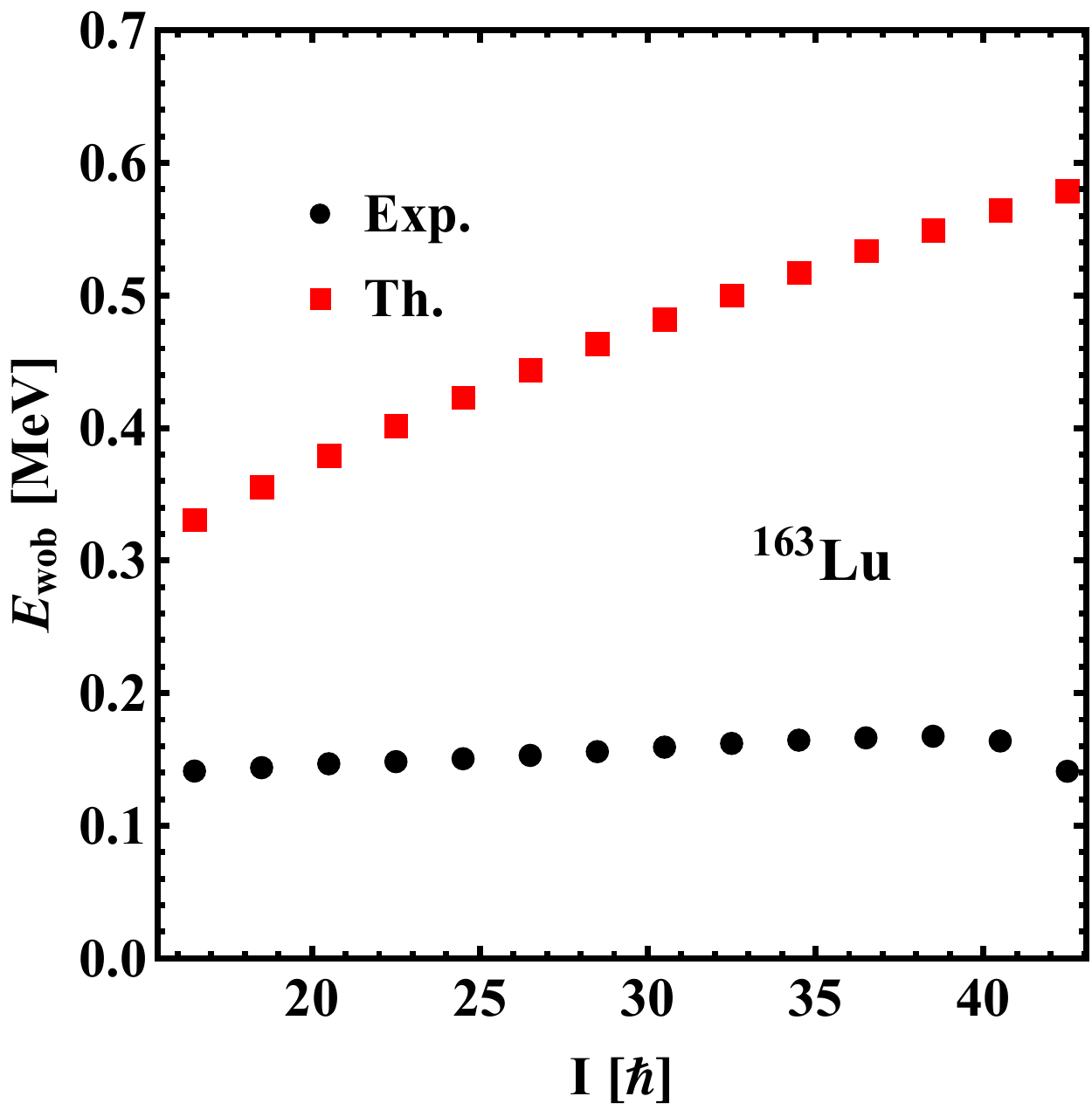}
{\scriptsize
\caption{(Color online)Wobbling energies, $E_{wobb}=E_1(I)-0.5(E_0(I+1)+E_0(I-1))$, with $E_1$ and $E_0$ defined as excitation energies  from TSD3 and TSD2 band, respectively. Experimental data are taken from Ref.\cite{Jens1}.}}
\label{Fig.2}
\end{figure}
The electric quadrupole intra- and inter-band transitions were calculated by using the transition operator
\begin{equation}
{\cal M}(E2,\mu)=\left[Q_0D^2_{\mu0}-Q_2(D^2_{\mu2}+D^2_{\mu-2})\right]
+e\sum_{\nu=-2}^{2}D^2_{\mu\nu}Y_{2\nu}r^2\equiv T^{coll}_{2\mu}+T^{sp}_{2\mu},
\end{equation}
with $Q_0$ and $Q_2$ taken as free parameters and the wave function:
\begin{eqnarray}
&&\Phi^{(1)}_{IjM}={\bf N}_{Ij}\sum_{K,\Omega}C_{IK}C_{J\Omega}|IMK\rangle |j\Omega\rangle\nonumber\\
&&\times \left\{1+\frac{i}{\sqrt{2}}\left[\left(\frac{K}{I}k+\frac{I-K}{k}\right)a^{\dagger}+
\left(\frac{\Omega}{j}k'+\frac{j-\Omega}{k'}\right)b^{\dagger}\right]\right\}|0\rangle_{I}.
\label{Phi}
\end{eqnarray}
with ${\bf N}_{Ij}$ standing for the normalization factor and $|0\rangle_{I}$ for the vacuum state of the bosons $a^{\dagger}$ and $b^{\dagger}$ determined by the classical coordinates $\varphi$, 
$\psi$ and the corresponding conjugate momenta $r$ and $t$ through the canonical parameters $k$ and $k'$. The expansion coefficient of the trial function corresponding to the minimum point, in terms of the normalized Wigner function, $C_{IK}$, were  analytically expressed in \cite{Rad018}. Since our fitting procedure  predicts that ${\cal I}_1$ is the maximal MoI, according for Ref.\cite{BMott}, $Q_0$ represents the quadrupole moment with respect to the one-axis while $Q_2$ is a measure of the asymmetry in the shape with respect to this axis. They determine the static moment and the B(E2) values for the intra-band transitions $I\to(I-2)$. Both of them are involved in the inter-band transitions $I\to (I-1)$ and $I\to (I+1)$.

Note that MoI's are free parameters, that is, no option for  their nature, rigid or hydrodynamic, is adopted. To be consistent with this picture the strengths  $Q_0$ and $Q_2$ were also considered as free parameters. However, this is not consistent with the structure of the single particle potential, which considers the collective quadrupole operator as emerging from the hydrodynamic model.
These are fixed by fitting the B(E2) values for one intra-band (TSD1) and one inter-band  ($TSD2\to TSD1$) transition. Thus, one obtained $Q_0=18.43eb$ and $Q_2=19.81eb$. The remaining B(E2) transitions and the quadrupole transition moments, listed in Tables II and III, are free of any adjustable parameter.
Results for the B(E2) values are compared with the corresponding data in Tables II and III.
\begin{table}
\begin{tabular}{|c|c|cc|cc|c|c|cc|cc|}
\hline
&&\multicolumn{2}{c|}{$B(E2;I^+\to (I-2)^+)$} &\multicolumn{2}{c|}{$Q_I$}&&&\multicolumn{2}{c|}{$B(E2;I^+\to (I-2)^+)$} &\multicolumn{2}{c|}{$Q_I$}\\
&&\multicolumn{2}{c|}{$[e^2b^2]$}    &\multicolumn{2}{c|}{$[b]$}&&&\multicolumn{2}{c|}{$[e^2b^2]$}    &\multicolumn{2}{c|}{$[b]$}  \\
\hline
TSD1&$I^{\pi}$&Th.  &  Exp.& Th. &Exp. &TSD2&$I^{\pi}$            &   Th.       &   Exp.   &  Th.   &  Exp.\\
\hline
&$\frac{41}{2}^+$&2.80&3.45$^{+0.80}_{-0.69}$&8.89&9.93$^{+1.14}_{-0.99}$&&$\frac{47}{2}^+$&2.71&2.56$^{+0.57}_{-0.44}$&8.71&8.51$^{+0.95}_{-0.73}$\\
&$\frac{45}{2}^+$&2.74&3.07$^{+0.48}_{-0.43}$&8.77&9.34$^{+0.72}_{-0.65}$&&$\frac{51}{2}^+$&2.66&2.67$^{+0.41}_{-0.33}$&8.62&8.67$^{+0.66}_{-0.53}$\\
&$\frac{49}{2}^+$&2.69&2.45$^{+0.28}_{-0.25}$&8.66&8.32$^{+0.47}_{-0.42}$&&$\frac{55}{2}^+$&2.62&2.81$^{+0.53}_{-0.41}$&8.53&8.88$^{+0.83}_{-0.64}$\\
&$\frac{53}{2}^+$&2.64&2.84$^{+0.24}_{-0.22}$&8.57&8.93$^{+0.38}_{-0.35}$&&$\frac{59}{2}^+$&2.58&2.19$^{+0.94}_{-0.65}$&8.46&7.82$^{+1.66}_{-1.15}$\\
&$\frac{57}{2}^+$&2.60&2.50$^{+0.32}_{-0.29}$&8.50&8.37$^{+0.54}_{-0.49}$&&$\frac{63}{2}^+$&2.54&2.25$^{+0.75}_{-0.48}$&8.39&7.91$^{+1.32}_{-0.84}$\\
&$\frac{61}{2}^+$&2.56&1.99$^{+0.26}_{-0.23}$&8.43&7.45$^{+0.49}_{-0.43}$&&$\frac{67}{2}^+$&2.51&1.60$^{+0.52}_{-0.37}$&8.34&6.66$^{+1.09}_{-0.76}$\\
&$\frac{65}{2}^+$&2.53&1.95$^{+0.44}_{-0.30}$&8.36&7.37$^{+0.82}_{-0.57}$&&$\frac{71}{2}^+$&2.49&1.61$^{+0.82}_{-0.49}$&8.28&6.68$^{+1.70}_{-1.02}$\\
&$\frac{69}{2}^+$&2.50&2.10$^{+0.80}_{-0.48}$&8.31&7.63$^{+1.46}_{-0.88}$ &&               &    &                      &         &                  \\
\hline
\end{tabular}
{\scriptsize
\caption{The E2 intra-band transitions $I\to (I-2)$ for TSD1 and TSD2 bands. Also, the transition quadrupole moments are given. Theoretical results (Th.) are compared with the corresponding experimental data (Exp.) taken from Ref. \cite{Gorg}. B(E2) values are given in units of $e^2b^2$,
while the quadrupole transition moment in $b$.}}
\label{Table 2}
\end{table}
\begin{table}
\begin{tabular}{|c|cc|cc|cc|}
\hline
&\multicolumn{2}{c|}{$B(E2;I^+\to (I-1)^+)$} &\multicolumn{2}{c|}{$B(M1;I^+\to (I-1)^+)$}&\multicolumn{2}{c|}{$\delta_{I\to(I-1)}$}\\
&\multicolumn{2}{c|}{$[e^2b^2]$}&\multicolumn{2}{c|}{$[\mu_N^2]$}&\multicolumn{2}{c|}{$[MeV.fm]$}  \\
$I^{\pi}$&Th.  &  Exp.& Th. &Exp. &Th.&Exp.\\
\hline
$\frac{47}{2}^+$&0.54&0.54$^{+0.13}_{-0.11}$&0.017&0.017$^{+0.006}_{-0.005}$&-1.55&-3.1$^{+0.36}_{-0.44}$\\
$\frac{51}{2}^+$&0.49&0.54$^{+0.09}_{-0.08}$&0.018&0.017$^{+0.005}_{-0.005}$&-1.58&-3.1$\pm 0.4$$^{a)}$\\
$\frac{55}{2}^+$&0.44&0.70$^{+0.18}_{-0.15}$&0.019&0.024$^{+0.008}_{-0.007}$&-1.61&-3.1$\pm 0.4$$^{a)}$\\
$\frac{59}{2}^+$&0.34&0.65$^{+0.34}_{-0.26}$&0.019&0.023$^{+0.013}_{-0.011}$&-1.64&-3.1$\pm 0.4$$^{a)}$\\
$\frac{63}{2}^+$&0.36&0.66$^{+0.29}_{-0.24}$&0.020&0.024$^{+0.012}_{-0.010}$&-1.66&\\
\hline
\end{tabular}
{\scriptsize
\caption{The B(E2) and B(M1) values for the transitions from TSD2 to TSD1. Mixing ratios are also mentioned. Theoretical results (Th.) are compared with the corresponding experimental (Exp.) data taken from Ref.\cite{Gorg}. Data labeled by $^{a)}$ are from Ref.\cite{Reich}.}}
\label{Tabel 3}
\end{table}

The magnetic transition operator used in our calculations is:
\begin{equation}
 {\cal M}(M1,\mu)=\sqrt{\frac{3}{4\pi}}\mu_N\sum_{\nu=0,\pm 1}\left[g_RR_{\nu}+qg_jj_{\nu}\right]D^1_{\mu\nu}\equiv M^{coll}_{1\mu}+M^{sp}_{1\mu},
\end{equation}
with $R_{\nu}$ denoting the components of the core's angular momentum with the corresponding gyromagnetic factor, $g_R=Z/A$, while $g_j$ is the free gyromagnetic factor for the single proton angular momentum $j(=13/2)$, which was quenched by a factor q=0.43 in order to account for the polarization effects not included in $g_j$. 

This factor takes care of the interaction of the odd-proton orbit with the currents distributed inside the core as well as of the internal structure of the proton, which may also influence its magnetic moment. To evaluate the transition matrix elements, the  involved states are written in the form:
\begin{equation}
\Psi_{IM}=\frac{1}{\sqrt{2j+1}}\sum_{M_R \Omega} C^{RjI}_{M_R \Omega M}C_{RK}|RMK\rangle|j\Omega\rangle .
\end{equation}
The expansion coefficients of the core's wave function  in the basis of the normalized Wigner function are denoted by $C_{RK}$.
Results for the relevant $B(M1)$ values of the inter-band transitions as well as for the mixing rations are collected in Table III.

Concluding the above analysis, the present formalism describes in a realistic fashion the experimental excitation energies in the bands TSD1, TSD2, TSD3, TSD4,   the intra- and inter-band B(E2) values, the transition quadrupole moments,  the dipole magnetic transitions from the levels of TSD2 to those of TSD1, B(M1), as well as the mixing ratios, $\delta_{I\to I-1}$ . 

Now, it is worth summarizing the specific features of the present approach: {\it i) the TSD2 band consists of the ground states provided by the variational principle of minimum action applied  for each angular momentum of the set I=R+j with R=1,3,5,..., and j=13/2; ii) the TSD3  states, I, are obtained by acting with the phonon operator on the TSD2 states of angular momenta I-1; 
iii) the negative parity band TSD4 is formed of the ground states corresponding to  I=R+j with R=1,3,5,... and j=9/2.}

In what follows we shall spend few words about the phase diagram associated with the classical energy function ${\cal H}$, for a given total angular momentum. From the equations of motion written in the Hamilton canonical form it results that the angles play the role of the classical coordinate, while  the variables $r$ and $t$ of the corresponding conjugate momenta. In virtue of this we may denote, more suggestively, the conjugate coordinates as:
\begin{equation}
q_1=\varphi,\; q_2=\psi,\; p_1=r,\;p_2=t.
\end{equation}
The critical manifolds associated to the classical energy function are determined from the equation:
\begin{equation}
\det \left(\frac{\partial ^2{\cal H}}{\partial (q_i)^{k}\partial (p_j)^{l}}\right)=0,\;i,j,=1,2; k,l=0,1,2;k+l=2.
\end{equation}
After some algebraic manipulations on the above equation one arrives at the equation:
\begin{equation}
C=0,
\label{Ceq0}
\end{equation}
where C has the expression from Eq.(\ref{BandC}). From (\ref{ecOm}) it is obvious that for this value of C, one solution vanishes. Thus, Eq.(\ref{Ceq0}) defines a Golstone mode which suggests a transition to a new nuclear phase. Eq.(\ref{Ceq0}) splits to the following equations:
\begin{eqnarray}
\hspace{-0.7cm}z&=&\frac{\left(1-4(I-j)^2\right)x^2+\left(4I^2+4j^2-8Ij+2j+2I-2\right)x-(2I+2j-1)}{G_1\left((2I-2j-1)x-(2I-1)\right)}\equiv f_1(x),\nonumber\\
\hspace{-0.7cm}z&=&\frac{\left(1-4(I-j)^2\right)x^2+\left(1-2(I+j)\right)y^2+2\left(2(I-j)^2+(I+j)+1\right)xy}{G_2\left((2I-2j-1)x-(2I-1)y\right)}\equiv f_2(x,y).
\label{f1andf2}
\end{eqnarray}
Here the following notations were used:
\begin{eqnarray}
x&=&\frac{A_1}{A_3},\;\;y=\frac{A_2}{A_3},\;\;z=\frac{V}{A_3},\nonumber\\
G_1&=&\frac{2j-1}{j(j+1)}\sqrt{3}\left(\sqrt{3}\cos \gamma+\sin\gamma\right),\;\;G_2=\frac{2j-1}{j(j+1)}2\sqrt{3}\sin\gamma.
\label{xandy}
\end{eqnarray}
For a fixed $\gamma (=17^{o})$, the equations (\ref{f1andf2}) represent two singular surfaces, having the asymptotic planes:
\begin{equation}
x=\frac{2I-1}{2I-2j-1},\;\;y=\frac{2I-2j-1}{2I-1}x.
\end{equation}

On the other hand, we recall \cite{Rad018} that the wobbling frequencies are obtainable by a quadratic  expansion of the energy function around the minimum point, which results in obtaining a Hamiltonian for two coupled oscillators. Quantizing the independent oscillators, the coupling term is diagonalized through a canonical transformation. Thus, the same frequencies as given by 
Eq.(\ref{ecOm}) are obtained. The frequencies for the uncoupled oscillators are real, provided the following restrictions hold:
\begin{eqnarray}
&&S_{Ij}A_1<A_2<A_3\;\; \rm{or}\;\;S_{Ij}A_1<A_3<A_2,\nonumber\\
&&A_3>T_{Ij}A_1+G_1,\;\;A_2>T_{Ij}A_1+G_2,
\label{ineq}
\end{eqnarray}
with
\begin{equation}
S_{Ij}=\frac{2I-1-2j}{2I-1},\;\;T_{Ij}=\frac{2j-1-2I}{2j-1}.
\end{equation}
The intervals (\ref{ineq}) together with the surfaces (\ref{f1andf2}) define, in the parameter space, sectors surrounded by separatrices which are conventionally called nuclear phases.
The phase diagram is presented pictorially in Fig.3, where the separatrices are shown. Therein, the planes $x=y,\;x=0,y=0$ are associated with the axial symmetric cases,  which are forbidden. 
\begin{figure}[ht!]
\includegraphics[width=0.8\textwidth]{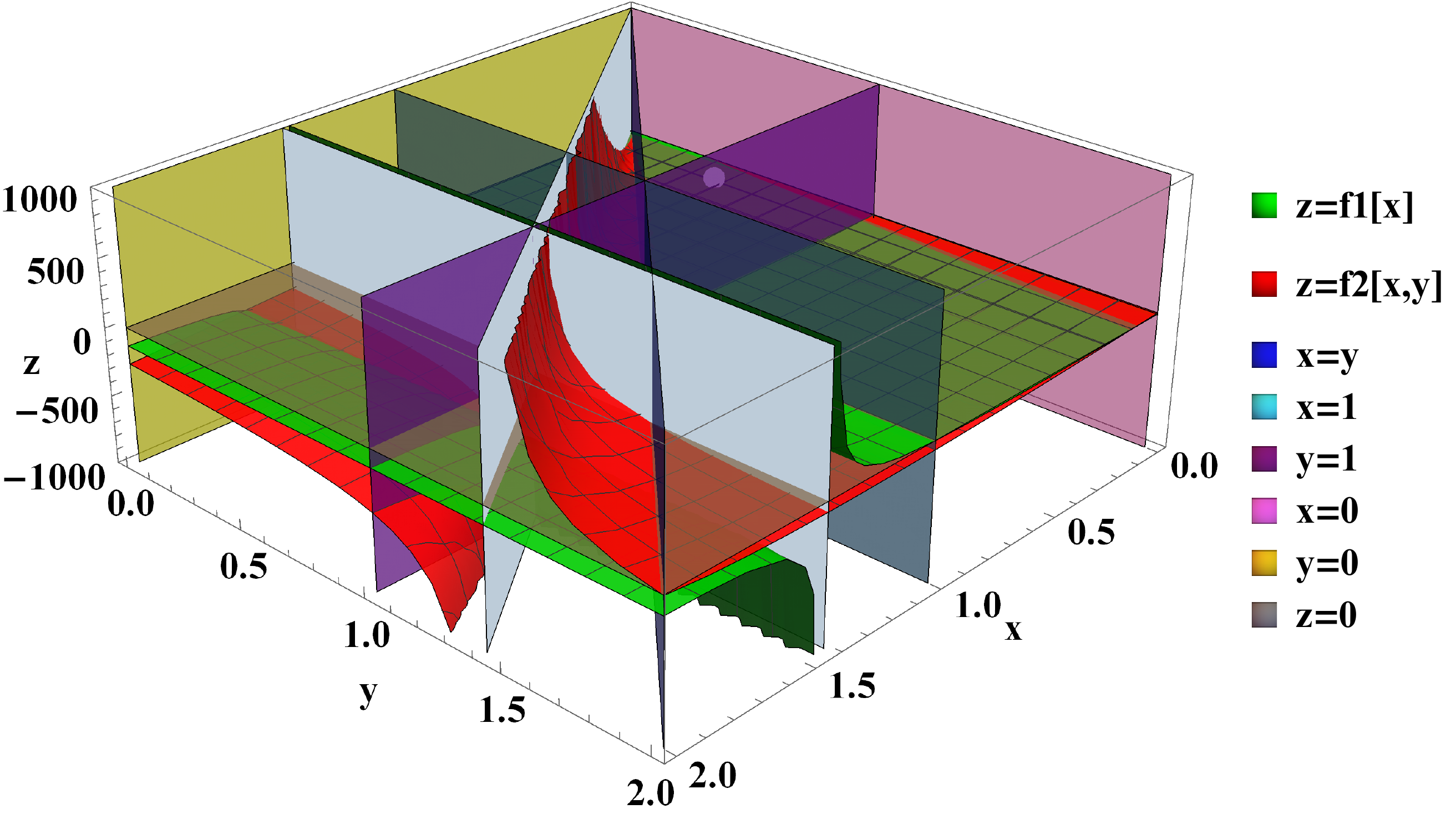}
\caption{(Color online)The phase diagram for a j-particle-triaxial rotor coupling Hamiltonian with j=13/2 and I=45/2.}
\label{Fig.3}
\end{figure}
The fixed MoI's and V are the coordinates of a point specified by a white small circle. Inside a given phase, the classical Hamiltonian has specific stationary points. If one of these is a minimum, then the classical trajectories surround it with a certain time period. If the point in the parameter space approaches the separatrices, the period tends to infinity \cite{Rad98}. When $V>0$, ${\bf j}$ is always oriented along the short axis, that is the axis x and the region where ${\cal I}_2>{\cal I}_1>{\cal I}_3$ is the phase where the transversal wobbling may take place. More specifically, this region is bounded by four surfaces: one being the diagonal plane, one is given by the second Eq. (22), the third one is the plane x=1, and the fourth which is defined by Eq. $z=f_2(x,y)$. For this region we have to depict the minimum of ${\cal H}$, if that exists. Furthermore, the frequencies describing the small oscillations around the minimum found are to be determined. Keeping MoI's inside the mentioned phase, we may proceed to fitting the experimental energies using the specific wobbling frequencies. Comparing the quality of the obtained fit with that of the present work, one may decide whether the transversal or longitudinal character of the wobbling prevails. It is worth noting that for $z<0$, the axis 3 ( the long one) is energetically favored in aligning ${\bf j}$. Therefore, another region where the transversal wobbling motion may show up, is bordered by the planes $x=0,\;y=1$, the asymptotic plane for the surface $z=f_1(x)$, and below  the surface $z=f_2(x,y)$. In the region between the two surfaces $z=f_1(x)$ and $z=f_2(x,y)$
the motion of the odd system is not allowed. Indeed, there $C<0$ and consequently the wobbling phonon frequencies become imaginary.

If in the expression of energy function one ignores the square root term, one obtains an energy surface having a minimum at the point $\varphi = \frac{\pi}{2},\;\psi=0,\; r= I,\; t=j$. The small oscillations around this minimum have the frequencies:

\begin{eqnarray}
&&\omega_1=(2I-1)\left[(A_3-A_2)(A_1-A_2)\right]^{1/2},\nonumber\\
&&\omega_2=(2j-1)\left[(A_3-A_1)+\frac{2\sqrt{3}V}{j(j+1)}\sin(\gamma+\frac{\pi}{3})\right]^{1/2}
\left[(A_2-A_1)+\frac{2\sqrt{3}V}{j(j+1)}\sin\gamma\right]^{1/2}.
\end{eqnarray}
Switching on the ignored term, the whole Hamiltonian is diagonalized through a canonical transformation of the RPA type, if that exists, the final frequencies being solutions of an equation of the type (\ref{ecOm}) with the coefficients $B$ and $C$ defined by:
{\scriptsize
\begin{eqnarray}
&-&B=(2I-1)^2(A_3-A_2)(A_1-A_2)+(2j-1)^2\left[(A_3-A_1)+\frac{2\sqrt{3}V}{j(j+1)}\sin(\gamma+\frac{\pi}{3})\right]
\left[(A_2-A_1)+\frac{2\sqrt{3}V}{j(j+1)}\sin\gamma\right],\nonumber\\
&&C=\left\{(2I-1)(2j-1)(A_1-A_2)\left[(A_2-A_1)+\frac{2\sqrt{3}V}{j(j+1)}\sin\gamma\right]\right\}\nonumber\\
&\times&\left\{(2I-1)(2j-1)(A_3-A_2)\left[(A_3-A_1)+\frac{2\sqrt{3}V}{j(j+1)}\sin(\gamma+\frac{\pi}{3})\right]-4IjA_3^2\right\}.\nonumber\\
\end{eqnarray}}
The existence conditions are:
\begin{equation}
x>y,\;\;y<1,\;\;z>\frac{G_2}{2j-1}(x-y),\;\;
z>\frac{4Ij}{(2I-1)G_1(1-y)}+\frac{2j-1}{G_1}(x-1).
\label{restr}
\end{equation}
The classical angular momentum components, corresponding to the minimum point, are:
\begin{equation}
I^{cl}_{1}=0,\;I^{cl}_{2}=-I,\;I^{cl}_{3}=0,\;j^{cl}_{1}=j,\;j^{cl}_{2}=0,\;j^{cl}_{3}=0.
\end{equation}
Such a situation is met with the hydrodynamic model for the MoI parameters and the particle-core potential given by Eq.(\ref{hassp}). The newly determined representation defines the true ground state which, however, might become unstable due to the Coriolis interaction. Such an instability reclaims a redefining of a new stable ground state which  
is associated with the longitudinal wobbling motion. Note that the transversal motion takes place if two severe conditions, marked by "{\it if exists}", are fulfilled.

For a rigid coupling, the coordinates $t, \psi$ disappear and the stationary point $(r=I,\varphi=\alpha)$, with $\alpha$ defined by
\begin{equation}
\cos\alpha=\frac{2j}{2I-1}\frac{A_1}{A_1-A_2},
\end{equation}
is a minimum of the energy function which results in having a stable ground state, but by ignoring the Coriolis interactions  determined by the core angular momentum components corresponding to the middle and long axes. Note that $\alpha\ne\frac{\pi}{2}$ and only for $I\gg j$, one may approximate 
$\alpha\approx \frac{\pi}{2}$. We may conclude that even for a rigid coupling of the odd proton along the short axis, the transversal wobbling mode may show up only in the limit of a very large I. Actually, the rigid coupling means that the initial Hamiltonian is truncated to a sum of two terms, one describing a triaxial rotor and one term linear in $I_1$, which cranks the system to rotate around the one-axis. When this happens, the longitudinal wobbling regime is achieved. Indeed, the first term favors the rotation around the middle axis, for hydro-dynamic MoI's, while the second term leads to a rotation around the  one-axis. The character of the wobbling motion is fixed by the result of the competition between the two effects.
Similarly, in the present formalism
the transversal wobbling appears with the price of ignoring important terms which leads to an energy function describing two independent oscillators. In this picture, the collective wobbling mode is determined exclusively by the core. Switching on the ignored interaction, new wobbling frequencies are obtained and the transversal picture is gradually blurred.

 It is conspicuous that the scenario presented here, points out that  the  picture where the transversal wobbling  shows up, corresponds to ideal restrictions \cite{Frau}, while within the Holstein-Primakoff description, the minimum for energy surface reflecting a transversal wobbling regime does not exist\cite{Tan017}, if one keeps all energy terms. Therefore, there is no contradiction between the two formalisms \cite{Frau018,Tana018}, since they deal with different Hamiltonians which, moreover, are subject to specific approximations.

{\it In conclusion, we can assert that the present formalism is based on the longitudinal concept and describes fairly well the main experimental properties of $^{163}$Lu.} 

{\bf Acknowledgment.} This work was supported by the Romanian Ministry of Research and Innovation through the project PN19060101/2019

\end{document}